\documentclass[letter]{aa} 
\usepackage{graphicx}
\usepackage{txfonts}
\usepackage{natbib}
\usepackage{psfig}
\bibpunct{(}{)}{;}{a}{}{,}
\begin{document}
\def\lsun{L_{\sun}}
\def\msun{M_{\sun}}

\title{The seeds of star formation in the filamentary infrared-dark cloud G011.11$-$0.12\thanks{{\em Herschel} is an ESA space observatory with science instruments provided by European-led Principal Investigator consortia and with important participation from NASA.  It is open for proposals for observing time from the worldwide astronomical community.}}


   \author{Th. Henning
          \and
          H. Linz 
          \and
          O. Krause
          \and
          S. Ragan
          \and
          H. Beuther
          \and
          R. Launhardt   
          \and
          M. Nielbock
          \and
          T. Vasyunina
          }

   \institute{Max-Planck-Institute for Astronomy, K\"onigstuhl 17, 69117
     Heidelberg, Germany \\
              \email{[henning, linz, krause, ragan, beuther, rl, nielbock, vasyunina]@mpia.de}
      }

\date{Received 31 March 2010 ; accepted 28 April 2010}

\abstract
{Infrared-dark clouds (IRDCs) are the precursors to massive stars and stellar clusters.  G011.11$-$0.12 is a well-studied filamentary IRDC, though, to date, the absence of far-infrared data with sufficient spatial resolution has limited the understanding of the structure and star-formation activity.}
{We use \it Herschel \rm to study the embedded population of young pre- and protostellar cores in this IRDC. }
{We examine the cloud structure, which appears in absorption at short wavelength and in emission at longer wavelength.  We derive the properties of the massive cores from the spectral energy distributions of bright far-infrared point sources detected with the PACS instrument aboard \it Herschel. \rm}
{We report on the detection and characterization of pre- and protostellar cores in a massive filamentary infrared-dark cloud G011.11$-$0.12 using PACS.  We characterize 18 cores directly associated with the filament, two of which have masses over 50~$\msun$, making them the best candidates to become massive stars in G011.11$-$0.12.  These cores are likely at various stages of protostar formation, showing elevated temperature ($<$T$> \sim$ 22~K) with respect to the ambient gas reservoir.  The core masses ($<$M$> \sim$ 24~$\msun$) are small compared to that in the cold filament.  The mean core separation is 0.9~pc, well in excess of the Jeans length in the filament.  }
{We confirm that star formation in IRDCs is underway and diverse, and IRDCs have the capability of forming massive stars and clusters.}


\maketitle
%

\section{Introduction}

Star formation occurs in cold, dense molecular cores.  In local low-mass star-forming regions, this process takes place in relatively isolated environments and is better understood compared to the distant regions in which massive stars form.  Specifically, it is the identification of the cold precursors to stars - the so-called ``pre-stellar cores'' that has led to a greater understanding of the initial conditions of low-mass star formation;  however, the high-mass counterpart has been more difficult to identify.  Thus it is not clearly understood how the early stages of massive star formation compare to that of low-mass stars.

Infrared-dark clouds (IRDCs) are complexes of cold (T $<$ 20~K), dense ($n > 10^4$~cm$^{-3}$) molecular gas, now believed to be the precursors to massive stars and star clusters \citep[e.g.][]{rathborne2006}, although there is diversity among this class of objects.  Early observations revealed that the cold dust in IRDCs absorbs the bright Galactic background emission at 8~$\mu$m \citep{egan_msx, hennebelle_isogal} and emits at sub-mm wavelengths \citep[e.g.][]{carey_submmIRDC}.  Later, high-resolution studies found that IRDCs exhibit significant substructure \citep[e.g.][]{ButlerTan2009, ragan_spitzer}, and (sub-)millimeter studies show that they host massive embedded star formation \citep[e.g.][]{Rathborne_2007_protostars, Beuther_protostars_IRDC, Hennemann2009}.  

The spectral energy distribution (SED) of massive embedded protostars is expected to peak in the far-infrared but is usually very faint in the mid-infrared.  The coldest cores are not readily detectable with {\em Spitzer} because of sensitivity limitations at 24~$\mu$m and confusion at longer wavelengths due to limited spatial resolution.  With {\em Herschel} \citep{A&ASpecialIssue-Herschel}, we now have the spatial resolution and the capability to perform large-area maps to search for these deeply embedded objects and to characterize the SEDs.

The large, filamentary IRDC G011.11$-$0.12 was selected as a target in the Early Phases of Star Formation (EPoS) {\em Herschel} key program (PI: O. Krause).  Together with complementary {\em Spitzer} data, we use PACS \citep{A&ASpecialIssue-PACS} and SPIRE \citep{A&ASpecialIssue-SPIRE} observations to constrain the properties of the massive embedded cores within this IRDC.

\begin{figure*}
\centering
\includegraphics[scale=0.47]{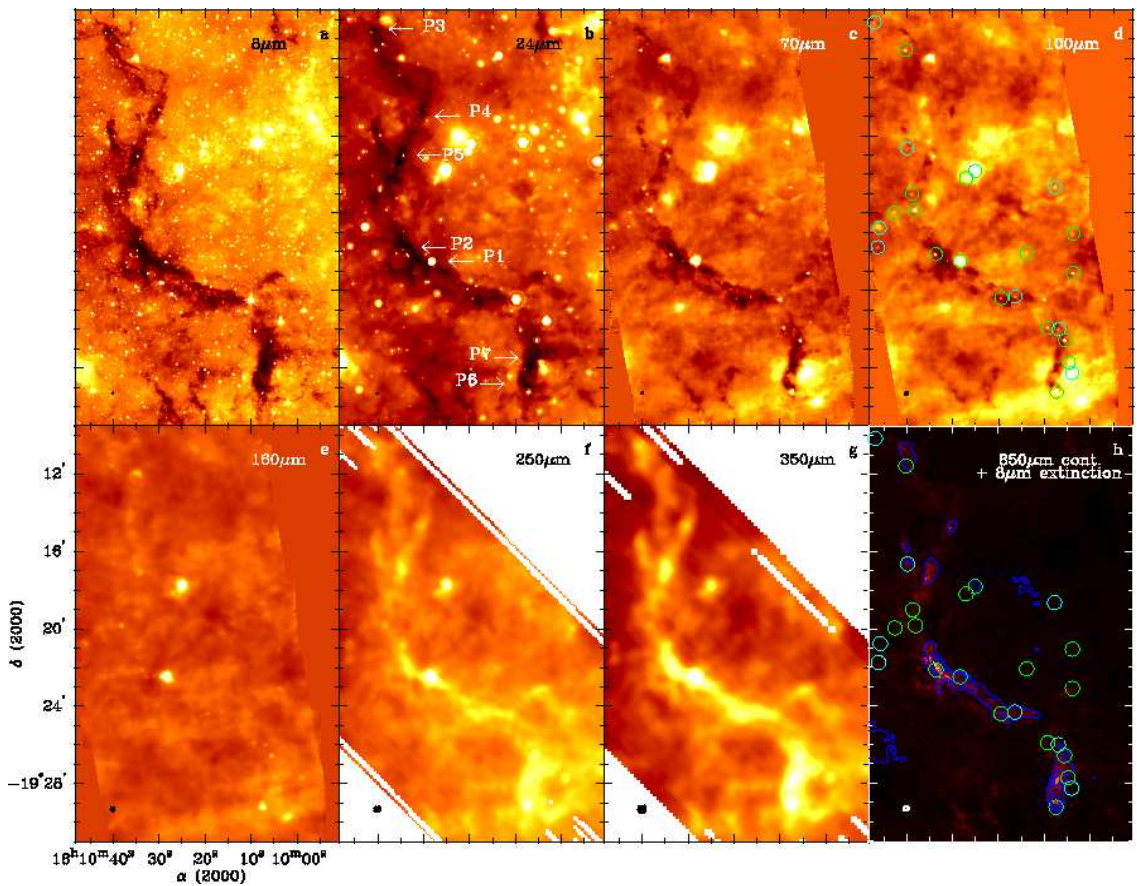}
\caption{The IRDC G011.11$-$0.12 from {\em Spitzer} to SCUBA wavelengths.  (a) {\em Spitzer} IRAC 8$\mu$m image.  (b) {\em Spitzer} MIPS 24~$\mu$m image.  The peak SCUBA positions (P1, P2,...) as assigned by \citet{Johnstone_G11} are indicated.  (c) {\em Herschel} PACS 70~$\mu$m image.  (d) {\em Herschel} PACS 100~$\mu$m image with the positions of analyzed sources (see Sect. 4) indicated.  Sources with no 24~$\mu$m counterpart are circled in cyan, sources with 24~$\mu$m counterparts are circled in green.  (e) {\em Herschel} PACS 160~$\mu$m image.  (f) {\em Herschel} SPIRE 250~$\mu$m image.  (g) {\em Herschel} SPIRE 350~$\mu$m image.  (h) 8~$\mu$m extinction map (derived from {\em Spitzer} data) with (blue) SCUBA 850~$\mu$m continuum contours (levels are 0.4 0.8 1.2 Jy beam$^{-1}$).  Source labeling same as in panel (d).  The approximate beam size is indicated in the lower-left corner of each panel.}
\label{fig:images}
\end{figure*}

Due to its high absorbing contrast against the Galactic background emission (particularly at 8~$\mu$m), the IRDC G011.11$-$0.12 ($d$~=~3.6~kpc) has been subject to detailed studies.  \citet{carey_msx} confirm the presence of dense molecular gas based on H$_2$CO observations.  \citet{carey_submmIRDC} report 850 and 450~$\mu$m continuum observations of the thermal dust emission which precisely follows the structure seen in absorption in the mid-infrared, and the column density reaches up to 10$^{23}$~cm$^{-2}$ \citep[see also][]{Parsons2009}.  The so-called P1 position is seen in emission in both 8 and 850~$\mu$m images (Fig. 1), a likely indicator of embedded star formation.  

\citet{Pillai_ammonia} observe G011.11$-$0.12 as part of an ammonia survey, including the (1,1) and (2,2) inversion transitions with 40$\arcsec$ spatial resolution.  The masses of four sub-regions (P1, P2, P6, and P7 in the \citet{Johnstone_G11} nomenclature) were found to be several hundred $\msun$ each, though other mass-tracing techniques find higher values for the total filament mass \citep[e.g.][]{Marshall2009}.  The NH$_3$ observations confirm that temperatures are very low throughout the filament, averaging 12~K.   In follow-up studies, observations of masers \citep{Pillai_G11}, extended excess 4.5~$\mu$m emission \citep[a tracer of shocked H$_2$][]{Cyganowski2008}, and molecular line emission \citep{Pillai_G11, Chen_infallEGO} have all supported the presence of ongoing star formation within the filament.  

\section{{\em Herschel} observations and data analysis}
\label{sec:obs}

The IRDC complex G011.11$-$0.12 was observed with the PACS instrument \citep{A&ASpecialIssue-PACS} on board of the {\em Herschel} Space Observatory \citep{A&ASpecialIssue-Herschel} on 2009 October 9, within the science demonstration program.  Scan maps in two orthogonal directions were obtained with the medium scan speed of 20$''$/s. Customized scan leg lengths (600$''$, 1200$''$) and scan position angles (12.5$^\circ$, 102.5$^\circ$) were employed in order to adapt for the elongated shape and orientation of the science object. In total, 2280, 2280, and 4560 seconds were spent for scanning on target in the blue (70 $\mu$m), green (100 $\mu$m), and red (160 $\mu$m) PACS filters, respectively. The raw data were reduced with the HIPE software \citep{HIPE}, version 3.0, build 455. Besides the standard steps leading to level-1 calibrated data, a second-level de-glitching as well as a correction for offsets in the detector sub-matrices were performed. Finally, the data were highpass-filtered using a large median window of 600$''$, in order to remove the effects of bolometer sensitivity drifts and 1/f noise along the course of the data acquisition.

Since we are mainly interested in the population of unresolved point sources, we used the IDL Starfinder program \citep{2000SPIE.4007..879D} to identify point sources and apply PSF photometry to the PACS data.  Reference PSFs (taken on the asteroid Vesta) were provided for all 3 filters by the PACS instrument team.

Maps at 250, 350, and 500 $\mu$m were obtained with SPIRE \citep{A&ASpecialIssue-SPIRE} on 19 October 2009. Two 720$''$ scan legs were used to cover the source. Four repetitions resulted in 340~s of scanning time with the nominal speed of 30$''$/s. The data were processed within HIPE \citep{HIPE} with the standard photometer script. Since for these observations no cross-scan data were secured, the resulting maps still show residual stripes along the one scan direction. The iterative de-striping algorithm devised by \citet{A&ASpecialIssue-Bendo} was invoked to mitigate this effect.

%

\section{Results and discussion}
\label{sec:results}

In {\em Spitzer} images, G011.11$-$0.12 appears mostly dark at both 8 and 24~$\mu$m, as shown in Fig.~\ref{fig:images}, though there are some indications of point sources at 24~$\mu$m.  However, the nature of these objects remains undefined because they lack counterparts at shorter wavelengths.  The IRDC filament is dark in the PACS 70 and 100~$\mu$m filter images, but shows little evidence of absorption at 160~$\mu$m.  In the continuum maps, starting with 250~$\mu$m SPIRE map to the SCUBA 850~$\mu$m map \citep{SCUBA_legacy}, shown in contours in Fig.~\ref{fig:images}, the extinction filament seen at shorter wavelengths now appears in emission. 

\begin{table}
\begin{minipage}{\columnwidth}
\caption{Photometric Properties of Protostars}
\begin{tabular}{lcccccl}
\hline \hline
Src. & RA & Dec. & Temp. & Lum. & Mass  & Notes\\
Num. & (J2000) &  (J2000) & (K) & (L$_{\odot}$) & (M$_{\odot}$) & \\
\hline
1 & 18:10:47.0 & -19:10:09 &    21 &     6 &     3 &    \\ 
2 & 18:10:40.3 & -19:11:34 &    23 &    26 &    9 & P3, e   \\ 
3 & 18:10:39.9 & -19:16:39 &    18 &     5 &     5 &    \\ 
4 & 18:10:38.8 & -19:19:01 &    24 &     8 &     1 &  e  \\ 
5 & 18:10:38.2 & -19:19:50 &    25 &    66 &     10 &  e  \\ 
6 & 18:10:45.9 & -19:20:46 &    16 &    15 &    40 &    \\ 
7 & 18:10:46.3 & -19:21:46 &    20 &     6 &     4 &    \\ 
8 & 18:10:33.7 & -19:22:08 &    21 &    25 &     9 &  P2, e  \\ 
9 & 18:10:28.4 & -19:22:29 &    24 &  1346 &   240&  P1, e  \\ 
10 & 18:10:03.8 & -19:23:06 &    23 &    27 &     7 &  e  \\ 
11 & 18:10:16.4 & -19:24:19 &    19 &    9 &     7 &    \\ 
12 & 18:10:19.4 & -19:24:24 &    21 &    11 &     5 &  e  \\ 
13 & 18:10:09.2 & -19:25:55 &    26 &    23 &     2 & P2, e   \\ 
14 & 18:10:06.9 & -19:26:00 &    20 &    9 &     5 &    \\ 
15 & 18:10:05.5 & -19:26:34 &    24 &    58 &     11 &   P7, e \\ 
16 & 18:10:04.7 & -19:27:43 &    24 &    25 &     5 &  P6, e  \\ 
17 & 18:10:04.0 & -19:28:16 &    23 &    28 &     6 &   P6 \\ 
18 & 18:10:07.4 & -19:29:15 &    23 &   139 &    82 & P6, e   \\ 
\hline

19 & 18:10:25.1 & -19:17:49 &    24 &   478&    79 &  e  \\ 
20 & 18:10:27.1 & -19:18:12 &    23 &    45 &     11 &  e  \\ 
21 & 18:10:07.7 & -19:18:39 &    21&    19 &     9 &    \\ 
22 & 18:10:42.7 & -19:19:58 &    21 &    11 &     5 &  e  \\ 
23 & 18:10:03.8 & -19:21:04 &    20 &     8 &     5 &  e  \\ 
24 & 18:10:13.8 & -19:22:04 &    17 &     9 &    15 &   e \\ 
\hline
\end{tabular}
Notes -- ``P'' labels indicate spatial correspondence to regions identified by \citet{Johnstone_G11}.  ``e'' indicates where we infer the presence of an embedded protostar based on a 24~$\mu$m detection.  Objects above the horizontal line are located ``on'' the IRDC filament; sources below are ``off.''
\label{phot}
\end{minipage}
\end{table}


\subsection{Characterization of embedded cores}

{\em Herschel} provides both the ability to make large-scale maps and the spatial resolution to isolate point sources from bright background emission for the first time.   At the given scales between 18000 and 43000~AU (our resolution limit from the 5 to 12$''$ PACS beam), we probe at the size scale of ``cores,'' as defined by \citet{BerginTafalla_ARAA2007}.   We detect cores seen as weak 24~$\mu$m point sources and new cores with no 24~$\mu$m counterparts (see Fig. 1).   The point sources follow the general structure of the extinction/emission maps quite well.  However, there are a few objects associated with bright emission to the north and west of the extinction map which are potentially associated with a more evolved stellar cluster.   

The PACS observations cover a crucial portion of the spectral regime in which cold cores are luminous.  For point sources with detections in the PACs bands (70, 100, and 160~$\mu$m), we fit the SED with modified Planck blackbody functions accounting for the wavelength dependence of optical depth assuming \citet{draine_lee} dust composition.  Note that each wavelength has a different beam size, thus the flux included in our photometric measurements is taken over correspondingly different spatial scales.  We do not include the 24~$\mu$m data point because the emission becomes optically thick, negating the assumptions of the optically thin blackbody fit.  The SPIRE data are important to constrain the large-scale properties of the filament, however with the poorer spatial resolution, we do not include them in the blackbody fits because the PACS cores cannot be identified.  The SCUBA data trace the colder surrounding filament, for which the peaks are not necessarily coincident with PACS point sources, so these data are also excluded from the SED fit.  

In Table~\ref{phot}, we present the position, model temperature, luminosity, and mass from the single-component blackbody model of the point sources with detections in all three PACS bands.  The mean mass, luminosity, and temperatures of the sources are 24~$\msun$, 96~$\lsun$, and 22~K, respectively.  Assuming fiducial uncertainties for the PACS band fluxes (10\%, 20\%, and 20\% for 70, 100, and 160~$\mu$m respectively), the temperature, luminosity, and mass values presented are accurate to approximately 4\%, 15\%, and 40\%, respectively.  We also note in Table~\ref{phot} which sources have 24~$\mu$m detections (16/24).  

Cores with a 24~$\mu$m detection likely host a protostar.  Cores with no 24~$\mu$m detection may be prestellar, but geometry may limit detectability at 24~$\mu$m (e.g. obscuration by dusty disk).  There are an additional 20 point sources not reported here detected in only two of the PACS bands that are also candidate prestellar cores.  Sources with no 24~$\mu$m detection (and even those with no 70~$\mu$m counterpart) cannot be definitively classified as prestellar.  \citet{XChen2010} has shown that (in a low-mass core) the absence of 70~$\mu$m emission cannot rule out the presence of a protostar.  Further observations (e.g. outflow signatures) are needed to verify the absence of deeply embedded protostars.  Perhaps the best candidates for truly "prestellar" objects are  the SCUBA peaks without any PACS emission.  These are column density peaks at an even more primitive stage, however without interferometric follow up at comparable resolution to the {\em Herschel} observations, the substructure of the prestellar material can not be fully characterized.  For example, \citet{Hennemann2009} and \citet{BeutherHenning2009} have shown that such dark regions in IRDCs fragment into a few massive cores.

In all, 18 of the 24 cores (75\%) are located ``on'' the IRDC, which we define as positions detected in the SCUBA map or on the 8~$\mu$m extinction map (see Fig. 1) in regions where SCUBA data do not exist.   There are two protostellar cores on the filament which have masses larger than 50~$\msun$, making them excellent candidates to host massive protostars.  We detect the P1 source (9 in Table 1) at all wavelengths, and we estimate the core mass to be 240~$\msun$, the highest in our sample.  

\begin{figure}
\hbox{
\vspace{1.0cm}
\hspace{-0.1cm}
\psfig{figure=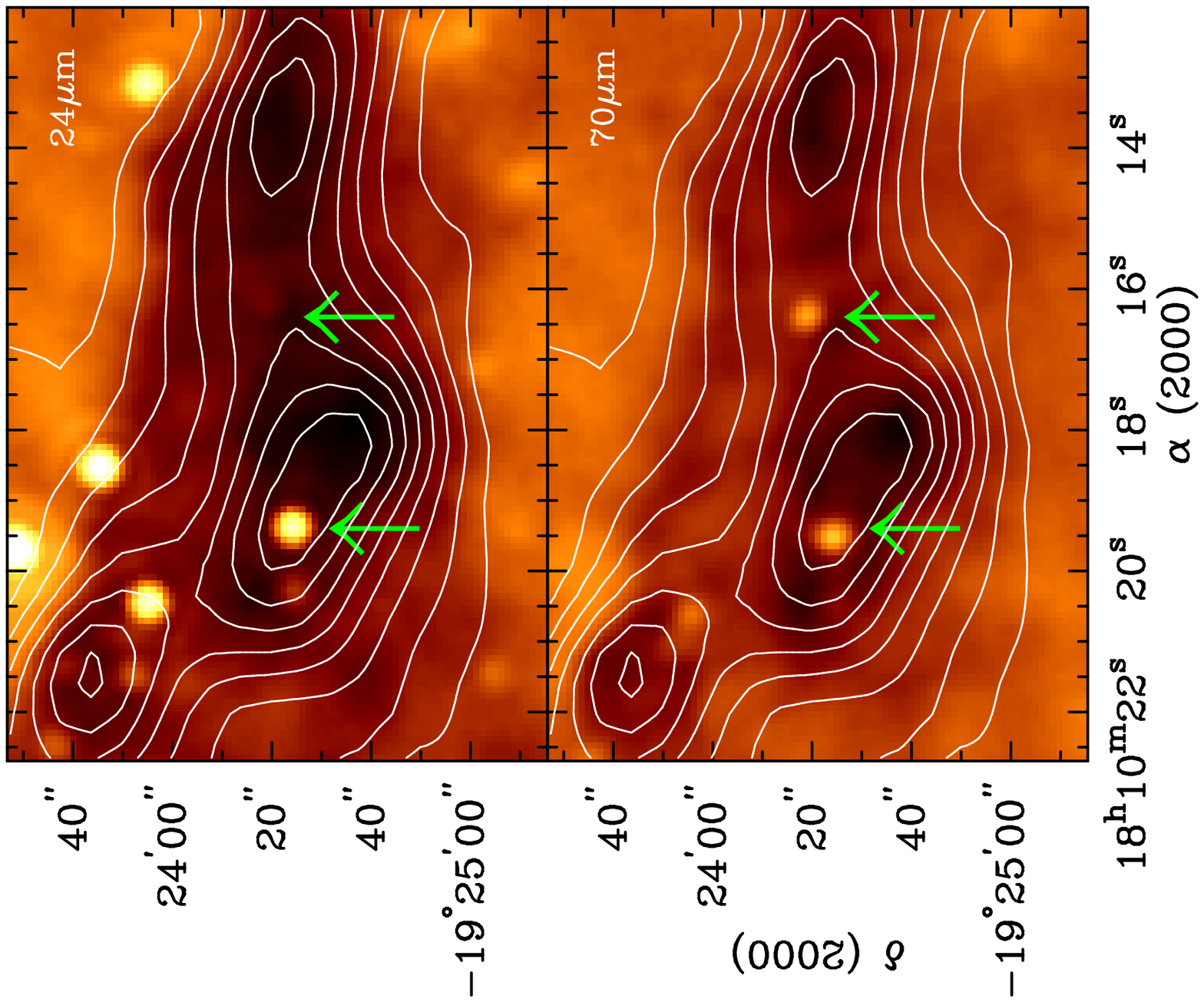,angle=270,height=5.0cm}
\hspace{0.05cm}
\psfig{figure=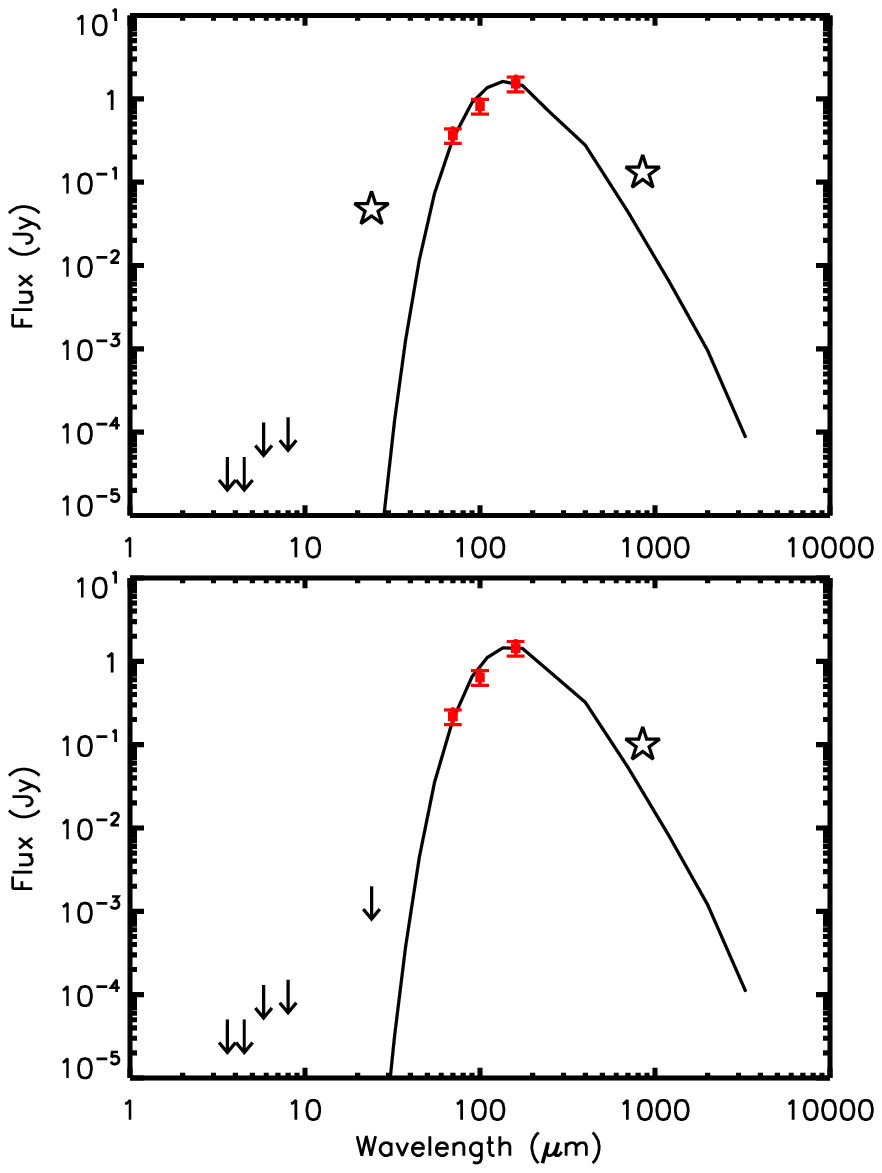,height=5.0cm,width=8.0cm}
}
\vspace{-1.0cm}
\caption{\textbf{Left:} SCUBA contours (levels: 0.1, 0.2, 0.3, 0.4, 0.5, 0.6, 0.7, 0.8, 0.9 Jy beam$^{-1}$) plotted over MIPS 24~$\mu$m image (top) and PACS 70~$\mu$m image (bottom).  Arrows indicate the location of point sources 11 (east) and 12 (west) from Table~1. \textbf{Right:} Spectral energy distributions of sources 11 (bottom) and 12 (top) detected within the extincted region. }
\label{fig:sed}
\end{figure}

The large scale map provides much information about individual parts of the filament, but now with high-resolution maps, we explore some detailed properties of subregions.  As an example, we discuss a region located south and west of P1, which includes sources 11 and 12.  In Fig.~\ref{fig:sed}, we show the 24~$\mu$m and 70~$\mu$m image of this small region spanning 1.7~pc, and the SEDs of the two sources.  The masses of cores 11 and 12 are 7 and 5~$\msun$, respectively, and the total mass reservoir estimated in the region is roughly 700~$\msun$ (derived from SCUBA 850~$\mu$m dust continuum).  The western source (11, bottom plot) is not detected at 24~$\mu$m (a pre-/protostellar core), and the eastern source (12, top plot) has a 24~$\mu$m detection in excess of the SED fit to the PACS data (a protostellar core).  While both in early evolutionary stages, we infer that source 12 is more evolved than source 11, even though it is likely that they originate from the same natal clump.  We also find that the local SCUBA emission peak does not directly coincide with either point source (see Fig.~\ref{fig:sed}), with the nearest source (11) offset by 19$''$ from the peak.  The core masses are small compared to the local clump mass within the filament.  Interferometric follow-up is key in determining the substructure within the region and the full census of pre- and protostellar objects \citep[cf.][]{Hennemann2009}. 

The SEDs presented in Fig.~\ref{fig:sed} include the PACS data used in the blackbody fit, limits on the {\em Spitzer} flux at 3.6, 4.5, 5.8, 8.0, and 24~$\mu$m, except in the case of core 12 where there is a 24~$\mu$m detection.  The data point at 850~$\mu$m on the SED is the flux estimated within a 23$''$ SCUBA beam \citep{SCUBA_legacy} at the position of the point source (corresponding to 83000 AU).  In all cases where there are SCUBA data, the flux within the SCUBA beam is in excess of the SED predicted from the blackbody fit to the PACS data.  This is not surprising because of the larger aperture and that the SCUBA measurement probes the emission from the colder filament, which includes only a small contribution from the warmer ($\sim$23~K) core.  Assuming a filament temperature of 12~K \citep{Pillai_ammonia} and $\kappa_{850}$ = 0.01 cm$^2$g$^{-1}$ \citep{draine_lee},  the flux within the SCUBA beam corresponds to a mass 40 - 80 times the core mass.  The SCUBA beam probes the mass in the outer envelope of the cores.  We do not claim to constrain this envelope size, but rather provide a uniform benchmark against which we can compare the cores to their immediate surroundings.  

\subsection{The nature of star formation in IRDCs}

The small aperture of {\em Herschel} allows us to probe the embedded population on the ``core'' scale in G011.11$-$0.12.  The mean core separation is 0.9~pc, which is many times the Jeans length, R$_J$ (for T~=~12~K  and 10$^4$ -- 10$^5$ cm$^{-3}$, $R_J$ = 0.06 to 0.2~pc), indicating that Jeans fragmentation seems not to be dominant in this IRDC.  Cores with a 24~$\mu$m detection are likely protostellar cores, and cores with only PACS detections can be either prestellar or protostellar (``pre-/protostellar'').  The SED fits give core masses from 1 to 240~$\msun$, showing that the core population, and therefore the star formation activity in this IRDC, is diverse.  The core temperatures are fairly homogeneous throughout the filament, averaging 22~K, which is $\sim$10~K warmer than the gas temperature measured in the large-scale ammonia maps.  This may indicate the presence of embedded star formation activity that would not be detectable with low-resolution observations \citep[e.g.][]{Pillai_ammonia}.  

The cores presented in this study represent a key early stage in protostar formation.  This important phase has until now been inaccessible, since {\em Herschel} allows large, high-resolution surveys of the far-infrared wavelength regime for the first time.  There are certainly also more primitive phases of star formation within IRDCs.  The filament is comprised of thousands of $\msun$ of cold gas, most of which is not in cores detectable with {\em Herschel} PACS.  The abundant cold gas may serve as a reservoir from which the cores accrete mass.  


\section{Conclusion}
\label{sec:conclusion}
%

{\em Herschel} enables us to perform large-scale, multi-wavelength surveys of star formation regions with unprecedented sensitivity and efficiency in the far-infrared.  In this contribution, we report the detection of embedded pre-/protostellar cores in the filamentary IRDC G011.11$-$0.12.  We characterize the SEDs of 24 cores.  The average mass, luminosity, and temperature of all the sources is 24~$\msun$, 96~$\lsun$, and 22~K.  Compared to the mass of the filament estimated from other surveys, there is an immense reservoir of mass that is not confined to pre-/protostellar objects.  The cores have an elevated temperature compared to that estimated from the \citet{Pillai_ammonia} large scale ammonia survey of the filament, which may be due to localized heating of the gas by protostars.

In forthcoming work, we intend to extend this investigation to perform more detailed radiative transfer modeling of the SEDs, connect the forming protostars to the kinematics, and extend these analyses to a larger sample of IRDCs observed as part of the EPoS program.  

\begin{acknowledgements}

PACS has been developed by a consortium of institutes led by MPE (Germany) and including UVIE (Austria); KUL, CSL, IMEC (Belgium); CEA, OAMP (France); MPIA (Germany); IFSI, OAP/AOT, OAA/CAISMI, LENS, SISSA (Italy); IAC (Spain). This development has been supported by the funding agencies BMVIT (Austria), ESA-PRODEX (Belgium), CEA/CNES (France), DLR (Germany), ASI (Italy), and CICT/MCT (Spain).  We thank the anonymous referee for useful comments that improved this letter.  

\end{acknowledgements}

\bibliographystyle{aa}
\bibliography{14635m}

\end{document}